\documentclass[
superscriptaddress,
 amsmath,amssymb,
 aps,
twocolumn,
]{revtex4-2}

\usepackage{graphicx}
\usepackage{dcolumn}
\usepackage{bm}
\usepackage{hyperref}
\usepackage{amsfonts}
\usepackage[bb=boondox]{mathalfa}
\usepackage{braket}
\usepackage{dsfont}
\usepackage{soul}
\usepackage[T1]{fontenc}
\usepackage[colorinlistoftodos,prependcaption,textsize=tiny]{todonotes} 

\usepackage{xcolor}

\usepackage{physics}
\usepackage{ulem}
\setstcolor{red}

\usepackage{color}
\usepackage{colortbl}
\usepackage{array}
\usepackage{multirow}

\usepackage{amsthm}

\begin{document}

\preprint{APS/123-QED}

\title{Adaptive Non-Gaussian Quantum State Engineering}

\author{Valerio Crescimanna}
\email{vcres052@uottawa.ca}
\affiliation{Blackett Laboratory, Department of Physics, Imperial College London, Prince Consort Rd, London, SW7 2AZ, United Kingdom}
\affiliation{Department of Physics, University of Ottawa, 25 Templeton Street, Ottawa, Ontario, Canada K1N 6N5}
\author{Shang Yu}
\affiliation{Blackett Laboratory, Department of Physics, Imperial College London, Prince Consort Rd, London, SW7 2AZ, United Kingdom}
\affiliation{Centre for Quantum Engineering, Science and Technology (QuEST), Imperial College London, Prince Consort Rd, London, SW7 2AZ, United Kingdom}
\author{Khabat Heshami}
\affiliation{Department of Physics, University of Ottawa, 25 Templeton Street, Ottawa, Ontario, Canada K1N 6N5}
\affiliation{National Research Council of Canada, 100 Sussex Drive, Ottawa, Ontario, Canada K1N 5A2}
\affiliation{Institute for Quantum Science and Technology, Department of Physics and Astronomy, University of Calgary, Alberta T2N 1N4, Canada}
\author{Raj B. Patel}
\email{raj.patel1@imperial.ac.uk}
\affiliation{Blackett Laboratory, Department of Physics, Imperial College London, Prince Consort Rd, London, SW7 2AZ, United Kingdom}
\affiliation{Centre for Quantum Engineering, Science and Technology (QuEST), Imperial College London, Prince Consort Rd, London, SW7 2AZ, United Kingdom}

\date{\today}

\begin{abstract}
Non-Gaussian quantum states of bosons are a key resource in quantum information science with applications ranging from quantum metrology to fault-tolerant quantum computation. Generation of photonic non-Gaussian resource states, such as Schrödinger's cat and Gottesman-Kitaev-Preskill states, is challenging. In this work, we go beyond existing passive architectures and explore a broad set of adaptive schemes. Our numerical results demonstrate a consistent improvement in the probability of success and fidelity of generating these non-Gaussian quantum states with equivalent resources. We also explore the effect of loss as the primary limiting factor and observe that adaptive schemes lead to more desirable outcomes in terms of overall probability of success and loss tolerance. Our work offers a versatile framework for non-Gaussian resource state generation with the potential to guide future experimental implementations.

\end{abstract}

\maketitle


\section{Introduction}\label{section:intro}




In quantum information science, the ability to engineer and manipulate quantum states is paramount for developing advanced quantum technologies. Non-Gaussian quantum state engineering, in particular, stands out as an essential task for many technologies, providing capabilities that extend beyond the limitations of Gaussian operations~\cite{lvovsky2020production,PhysRevLett.130.090602}. Gaussian states and operations, characterised by their ease of implementation and mathematical simplicity, form the backbone of many quantum information protocols~\cite{WANG20071,aaronson2011computational, RevModPhys.84.621,wagner2012continuous}. However, the intrinsic properties of Gaussian states are insufficient for achieving universal quantum computation and certain types of quantum error correction~\cite{PhysRevLett.102.120501, PhysRevA.81.010302,PhysRevA.98.022335,PhysRevA.97.032335}.

Non-Gaussian quantum states, which deviate from the Gaussian distribution in their Wigner function representation, offer unique and powerful resources necessary for the realization of more sophisticated quantum information tasks. These states enable the implementation of quantum gates and operations that are essential for universal quantum computation, facilitating complex quantum algorithms that cannot be accomplished with Gaussian states alone~\cite{doi:10.1080/09500340601101575, PhysRevA.93.022301}. 
In quantum communication, non-Gaussian quantum states are strong candidates for overcoming the limitations imposed by repeaters relying solely on Gaussian operations~\cite{PhysRevA.90.062316,rozpkedek2021quantum}.
Furthermore, non-Gaussian operations are pivotal in enhancing the robustness and efficacy of quantum error correction schemes, thereby improving the fidelity and scalability of quantum information systems~\cite{Terhal_2020}.

The engineering of non-Gaussian states is also crucial for quantum metrology and sensing~\cite{zurek2001sub,AGilchrist_2004, Duivenvoorden2017,Zhuang_2020,doi:10.1126/sciadv.adw9757}, where enhanced precision measurements are sought. By leveraging the distinct properties of non-Gaussian states, quantum sensors can achieve sensitivities that surpass those of Gaussian squeezed states.

Continuous-variable (CV) encoding of optical quantum information has been studied for many years. To achieve an advantage over classical computing, it is well known that some element of non-Gaussianity is required~\cite{Bartlett-2002a,Bartlett-2002b}. This may be introduced either in the resource states used in the protocol, in circuit operations, or at the measurement stage.

For quantum computing, discretizing and embedding quantum information in the infinite-dimensional space of an oscillator offers a route to implementing bosonic error-correcting codes (BECCs)~\cite{PhysRevX.10.011058,CAI202150,Joshi_2021,BRADY2024100496} for fault-tolerant quantum computing using far fewer physical resources compared to discrete-variable encoding alone. 

Nevertheless, fault-tolerant quantum computation can be achieved by encoding information into specific non-Gaussian quantum states capable of detecting and correcting shifts in the quadrature space due to noise in the system~\cite{PhysRevA.64.012310}. Efficient engineering of non-Gaussian states is therefore highly desirable to scale up fault-tolerant quantum computation (FTQC). However, while sources of non-Gaussian states can be deterministically produced with high fidelity in superconducting circuits, their generation presents greater challenges in photonic implementations where generation necessitates nonlinear gates or probabilistic preparation of the initial input states~\cite{PhysRevLett.103.203601,PRXQuantum.2.030204,Endo:23,PhysRevResearch.5.033156,CAI2023107171,9951256,Winnel2024}.

Here, we explore alternative conditional sources of optical non-Gaussian states. Specifically, we introduce an adaptive scheme comprising two or more layers of preparation, where the configuration of the subsequent circuit is set depending on the number of photons detected in the early stages. We compare their efficiency in terms of the quality of the generated state and the probability of success, for different input states and circuit configurations. Finally, we offer an outlook for potential experimental implementations.

\section{Approaches to non-Gaussian quantum state engineering with GBS devices}\label{section:GBS state engineering}
\subsection{Non-Adaptive state engineering}\label{section:GBS non-adaptive}

In the continuous-variable domain, an $n$-mode quantum state $\rho$ is Gaussian if its Wigner function  $W(\rho)$ is Gaussian in shape, that is if it can be expressed as~\cite{PRXQuantum.2.030204} 
\begin{equation}\label{eq:WignerGaussianState}
    W(\bm{x}) = \frac{\exp\left[\frac{1}{2}(\bm{x}-\bm{\xi})^\top\bm{V}^{-1}(\bm{x}-\bm{\xi})\right]}{(2\pi)^N\sqrt{\det\left(\bm{V}\right)}}
\end{equation}
where $\bm{V}$ is the covariance matrix and $\bm{\xi}$ is the displacement vector of the state.
Any quantum state whose Wigner function cannot be expressed in terms of \eqref{eq:WignerGaussianState} is non-Gaussian. 
Similarly, quantum operators, $\mathcal{O}$ are Gaussian if they transform one Gaussian state into another, and are non-Gaussian, otherwise.  

Considering photonic implementation, the operators that can be deterministically realized on the optical table—namely, single-mode squeezing, displacement, phase shifts, and two-mode beam splitting—are all Gaussian operations belonging to the Clifford group. Therefore, with these operations alone, it is impossible to transform the Gaussian vacuum state or any other Gaussian state into a single-mode non-Gaussian state.

\begin{figure}[!t]
\begin{center}
    \includegraphics[trim=90 90 90 60, clip,width=0.4\textwidth]{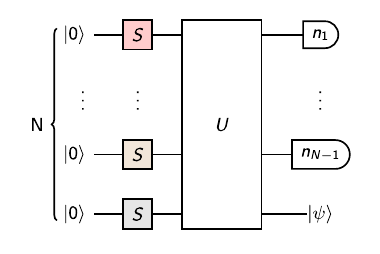}
\caption{GBS device with $N$ squeezed displaced input vacuum states and $N-1$ PNRDs.}
\label{fig:GBSdevicepure}
\end{center}
\end{figure}

A conditional scheme for non-Gaussian state generation with linear optics, in a Gaussian Boson sampling (GBS-like) device, was originally proposed by Su~\textbf{et al.}~\cite{PhysRevA.100.052301}. In this protocol, illustrated in Fig.~\ref{fig:GBSdevicepure}, non-Gaussianity is introduced by projection measurements on non-Gaussian Fock states in the ancillary modes. Starting with an $n$-mode Gaussian state, measuring predefined combinations of photons in $n-1$ modes can herald the desired non-Gaussian state if the parameters of the source are set properly. However, in general, even the optimal squeezing intensities and angles of the passive interferometer do not guarantee measuring the expected pattern. Consequently, the probability of detecting the expected pattern, i.e., success probability, is also taken into account alongside the distance of the generated state from the target when optimizing parameters and measurement patterns.

Numerical optimizations of the parameters in GBS-like sources have been performed to realize Schr\"{o}dinger cat states~\cite{PhysRevA.100.052301} and Gottesman-Kitaev-Preskill (GKP) states~\cite{PhysRevA.101.032315}. The results of these optimizations depend on the number of modes considered, the measurement patterns, and the optimization weight given to the quality of the output over the success probability. Overall, the results are promising but show significant room for improvement, especially in terms of probability, which is essential for enhancing computational efficiency.

Alternative schemes have indeed been proposed to improve the quality of the sources, for example, by incorporating additional non-Gaussian resources as input~\cite{PhysRevA.109.023717}.
Moreover, one can envisage adapting the source so that a portion of the circuit can be reconfigured if some measurements do not yield the expected outcome.

\subsection{Adaptive state engineering}\label{section:GBS adaptive}

The main limitation of a conditional source made with a GBS-like device lies in the probability of generating the state. Efficient state generation is crucial for most applications. Running the source several times would increase the operation time and, as a consequence, the probability of experiencing loss. On the other hand, multiplexing several sources requires a proportional use of costly resources.
Moreover, large-scale computations rely heavily on non-Gaussian resources; therefore, even marginal improvements in the success rate of the source can yield substantial benefits for the overall computational process.

The success probability depends on the parameters of the sources, namely, the number of modes, the beam splitter ratios and phase shifting angles, and the squeezing intensities. Although the parameters can be tuned to increase the probability of success, this is done at the expense of the quality of the state that is generated in the heralded mode. 
Such an argument becomes particularly evident when considering the number of modes the source uses. Some states can be reached only with a minimum number of modes, but the combined probability of measuring a given number of photons in each mode decreases as a consequence~\cite{PhysRevA.100.012326}.

Indeed, the system in~\ref{section:GBS non-adaptive} does not yield the desired state unless all measurement outcomes are as expected. 
Still, the measurement pattern that enables the generation of the target state is, in general, not unique: a high-quality state can be generated using different circuits, each employing different combination of measurements, provided that each circuit is tuned according to its corresponding array of detections. For instance, adding a swap gate between two modes in the unitary passive interferometer forms a new interferometer that adapts to a permutation of the original measurement pattern.

\begin{figure*}[t]
\centering
\includegraphics[trim=20 680 20 35, clip,page=2,width=1\linewidth]{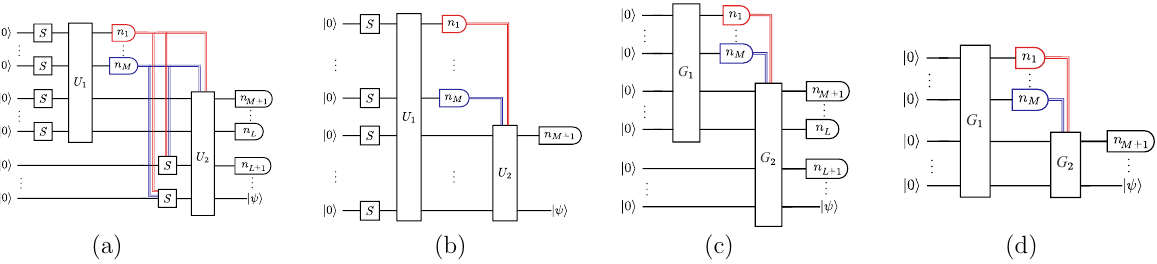}
\caption{\label{fig:AdaptiveSchemes} (a) Adaptive non-Gaussian state generation with GBS-like devices. Here, $S$ represents the squeezing operator, $U$ represents the unitary linear interferometers, and $n_i$ denotes the measurement outcome of the PNRD at the $i$-th mode. Single lines represent quantum channels, while double lines represent classical channels. (b) Adaptive scheme where the interferometer $U_2$ is the sole element in the adaptive stage. (c) Adaptive scheme analogous to (a) with general Gaussian operations instead of passive unitary interferometers. (d) General adaptive scheme with Gaussian operations.}
\end{figure*}

However, once a measurement pattern is committed to, the source can only achieve success with that specific permutation. This restriction stems from the fact that the measurement layer represents the final step, preventing any retroactive modifications to the interferometer or input states. 
This problem might be partially addressed by considering an alternative scheme, the adaptive GBS-like source. In this alternative source, depicted in Fig.~\ref{fig:AdaptiveSchemes},
 only some of the modes are manipulated in the first stage, and a subset of them gets measured. 
The measurement results obtained are then used to inform the second part of the circuit. The squeezing intensities of the remaining states, as well as the parameters of a second interferometer that acts on the unmeasured modes, are adjusted according to the outcomes of the first measurement.

A specific instance of this scheme is illustrated in Fig.~\ref{fig:AdaptiveSchemes},
where all modes initially pass through a universal interferometer in the first layer, while the second interferometer is the only element that adapts based on the measurement outcomes from the first subset of modes.

\begin{figure}[!t]
\centering
\includegraphics[trim=420 480 60 280, clip,,width=0.7\linewidth]{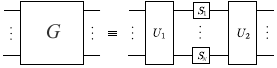}
\caption{\label{fig:BMdecomposition} Scheme of the Bloch-Messiah decomposition.}
\end{figure}

The adaptive source can be generalized by allowing inline squeezing of states besides vacuum. In this case, we consider using generic Gaussian operations 
$G$, defined in terms of symplectic matrices, instead of the unitary interferometers as depicted in Fig.~\ref{fig:AdaptiveSchemes}.
According to the Bloch-Messiah decomposition~\cite{PhysRevA.71.055801}, symplectic operations can be expressed as a series of two interferometers interspersed with a layer of inline squeezing single-mode operators.
A schematic of the Bloch-Messiah decomposition is shown in Fig.~\ref{fig:BMdecomposition}.

In this scenario, the most general scheme consists of a series of symplectic operations acting on all available modes at each step, illustrated in Fig.~\ref{fig:AdaptiveSchemes}.

A further generalization of the scheme can occur by considering a sequence of adaptable layers, with the information fed forward through each successive layer. In this regard, it is important to note that the number of modes and measured photons necessary to generate a state depends on the state itself and can be linked to its Stellar rank. Consequently, for states that are easier to generate, the initial layers of the adaptive scheme could be sufficient, and the additional layeers may come into play only when the measurement pattern in the first layer is not desired, to increase the overall probability of success.

It is worth stressing that also in the adaptive case the optimization of the parameters used to find the ideal source is conventional. It can be biased towards either the success probability or the fidelity of the output state. Moreover, in the adaptive scheme, the optimization can be defined in such a way to favor the generation of the state already at the early stages or to maximize the overall probability by fixing the total number of steps. Finally, if we anticipate a high cost for the adaptivity of the circuit, the optimization could be built to favor the success for a given measurement. 

The feed-forward scheme described so far can be enriched by introducing single Fock states into the input modes instead of vacuum states. Indeed, it has been shown in the non-adaptive scheme in Fig.~\ref{fig:GBSdevicepure} that introducing these non-Gaussian input states as an additional source of non-Gaussianity improves the efficiency of the GBS-like device in generating GKP and Schrödinger cat states. Here, we evaluate whether the Fock states actually enhances the efficiency of the adaptive source and compare its performance to a non-adaptive scheme with the same number of single Fock states as input. 
Schemes with different distributions of input Fock states are not considered here, but for them, we would anticipate that increasing the number of single-photon states would improve the overall efficiency of the source, similarly to what happens in non-adaptive sources~\cite{PhysRevA.109.023717}. In contrast, injecting an equivalent number of photons into a single mode would not be as effective, as suggested by the calculations on simulability via coherent state decomposition~\cite{Marshall:23}.
Finally, we observe that adaptivity, consisting of the information from the measurement outcomes of some mode being fed forward to modify the rest of the circuit, has been shown to be beneficial for other quantum resource state sources that rely on homodyne detection and passive interferometers seeded with non-Gaussian states~\cite{PhysRevA.110.012436}. However, these architectures cannot be as easily generalized to produce arbitrary target states as GBS-like sources.

\section{Photon loss in adaptive state engineering}\label{section:adaptive+loss}

The discussion developed so far does not impose any specific constraints on the physical implementation on which the source should be built, and experimentalists can choose their preferred one according to costs, availability, space, efficiency, and applications of the source itself. In any case, in the physical realization of the GBS-like device, certain physical limitations emerge that significantly impact the efficiency of the source. The primary issue encountered in photonic implementations in the real world is loss. For both the non-adaptive and adaptive sources introduced in section~\ref{section:GBS state engineering}, loss affects the circuit mainly at the input and output.

\begin{figure}[t]
\centering
\includegraphics[trim=100 220 180 380, clip,,width=0.9\linewidth]{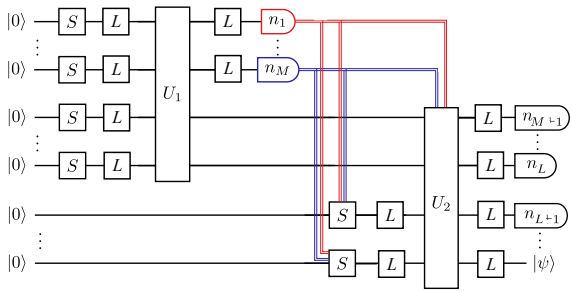}
\caption{\label{fig:AdaptivewLoss} Adaptive scheme with loss occurring before and after $U_{1,2}$.}
\end{figure}

A scheme including loss is illustrated in Fig.~\ref{fig:AdaptivewLoss}. Uniform losses are introduced in every mode twice: right after the squeezing operation acting on the vacuum state, and before detection in the herald modes and at the output of the unmeasured mode. The quality of the output state, and to a lesser extent the probability of detecting the expected measurement pattern, decrease depending on the intensity of the loss. The efficiency of the source's dependence on loss can thus serve as an additional figure of merit for comparing adaptive and non-adaptive schemes. Similarly, the performance of schemes using non-Gaussian input states can be evaluated. 

If the loss is known a priori, the circuits can be optimized accordingly. Simulations on non-adaptive schemes have already been performed~\cite{PhysRevA.101.032315}. Similarly, we evaluate how loss affects the results of an adaptive scheme when the parameters are optimized for it, and assess whether the robustness of this scheme, which relies on feedforward information, is comparable to that of the non-adaptive approach. 

\section{Method}\label{section:metohd}

In order to have a fair evaluation of the adaptive scheme introduced in Section~\ref{section:GBS adaptive}, we compare its performance with that achievable using an analogous non-adaptive scheme as described in Section~\ref{section:GBS non-adaptive}. 
In particular, we consider the amount of resources used in both schemes. We use the total number of modes in each source as the main criterion for a fair comparison. Whenever a squeezing bound is set in each mode, it is the same for both schemes. Similarly, the same number of single Fock states is used in each model, if any. 

The sources are optimized using classical simulations of the circuit implemented with the Python library `Mr Mustard' developed by Xanadu. Due to simulation constraints and for a clearer proof of principle, we consider small sources consisting of three and four modes, but we anticipate that the adaptive protocol would perform equally well or better as more modes can be modified based on the information fed forward. 

The source parameters are optimized to maximize a reward function that is a linear combination of the probability of success $P$ and fidelity with the target state $\mathcal{F}$. In particular, we set the reward function to $\mathcal{F} + P $, thus giving the same weight to both the figures of merit, as prioritizing one over the other may lead to sources that either have poor success rate or that generates low-quality states. To ensure a fair comparison between the protocols, the parameters of the adaptive source are optimized to match the fidelity achieved by the classical counterpart.
The chosen target states are the Schrödinger cat states and the GKP grid states, both prominent examples of non-Gaussian quantum states with numerous applications such as communication, cryptography, and computation~\cite{PhysRevA.65.042305, PhysRevA.68.042319, PhysRevLett.111.120501,  xu2023autonomous,PhysRevResearch.3.033118,PhysRevA.101.032315}. Specifically, we selected the amplitude and squeezing parameters of the cat states to be sufficiently challenging to produce using a two-mode circuit, representing the minimal non-adaptive approach, yet achievable with high fidelity using a three-mode source, which is the minimal configuration for a non-trivial adaptive scheme. 
Regarding GKP states, we approximated the ideal unphysical state $\ket{\bar{0}_I}$ with $\ket{\bar{0}_{\Delta}}$ such that
\begin{equation}\label{eq:CanonicalNormalization}
    \ket{\bar{0}_I}\rightarrow\ket{\bar{0}_{\Delta}}\propto\sum_{n=-\infty}^{\infty}e^{\frac{1}{2}\Delta^2(2n\sqrt{\pi})^2}\bar{X}^2\ket{\Delta}_q,
\end{equation}
where
\begin{equation}
    \braket{q}{\Delta}=\left(\frac{1}{\pi\Delta^2}\right)^\frac{1}{4}e^{-\frac{q^2}{2\Delta^2}}.
\end{equation}
Specifically, we target the truncated core state in
\begin{equation}\label{eq:GKP4}
   \ket{0_{A4}}=S(r)\underbrace{\sum_{n=0}^4 c_n\ket{n}}_{\text{Core state}}
\end{equation}
where $c_n$ are tuned to maximize $\expval{0_{\Delta=10\text{dB}}}{0_{A4}}$~\cite{PhysRevA.101.032315}.

\section{Results}

\subsection{Adaptive state engineering with squeezed states and PNRD}\label{section:ASE with squeezed states}
\subsubsection{Odd cat states}\label{section:cat state}
\begin{figure}[h] 

\centering 

\includegraphics[trim=120 310 350 290, clip,,width=1\linewidth]{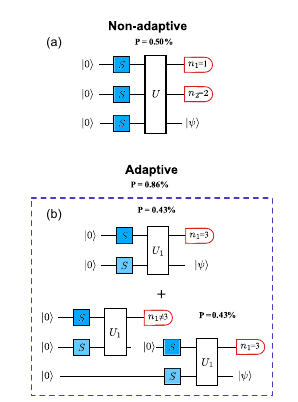} 

\caption{\label{fig:OddCatState} Comparison between the (a) non-adaptive and (b) pseudo-adaptive schemes for generating the odd cat state. The squeezing is bounded by $r_{\text{max}}=0.5$. The darker shades in the squeezing boxes correspond to higher squeezing intensities, with the darkest blue representing $r=0.5$. The pseudo-adaptive scheme is equivalent to running the non-adaptive scheme a second time. The target state is the odd cat state with $\alpha=\sqrt{6}$. The plus sign indicates that the probabilities corresponding to the two possible outcomes of the adaptive sources can be summed.} 
\end{figure}

First, we compare two schemes made of three modes as shown in Fig.~\ref{fig:OddCatState}. We set the maximum threshold of the squeezing intensity of the input states to $r = 0.5$.  The target state chosen for this comparison is a squeezed odd cat state with $\alpha=\sqrt{6}$ , and $r=0.5$, that, as desired, is a state that should present a sufficient challenge when produced with only two modes, yet it should still be attainable with satisfactory fidelity.  

Initially, the optimization of the adaptive scheme is made such that it favours the generation of the state after the first measurement. Indeed, a two-mode circuit can be interpreted as a three mode circuit in which the third mode is not interfered. In this case, the fidelity between the generated and the target state is the same for both the adaptive and non-adaptive scheme when $n_{\text{tot}}=3$ photons are measured: $\mathcal{F}\simeq97.6\%$. The probability, on the other hand, is $P\simeq0.43\%$ for the two-mode circuit and $P\simeq0.50\%$ for the three-mode circuit. Therefore, there is a relative gain in probability by adding a mode of approximately $16\%$. 
However, we observe that when the initial measurements $n_1$ of the adaptive circuit differ from the expected measurements, no combination of squeezing and interferometer $U_2$ can be found that produces a state close enough to the target. 
The scheme is effective only if we consider running it again, resulting in a doubling of the probability of producing the state, i.e., $P\simeq0.86\%\geq0.50\%$. This last result is indeed achievable with two beam splitters and two PNRDs, which are the same resources used in the non-adaptive scheme. Alternatively, one can view it as a scheme in which the state heralded by a measurement outcome different from the expected one is discarded, and a squeezed vacuum state is used instead to interfere with the state in the third mode at the second beam splitter. The schemes with their probabilities are shown in Fig.~\ref{fig:OddCatState}. 

\begin{figure}[t] 
\centering 
\includegraphics[trim=150 550 320 50, clip,width=1\linewidth]{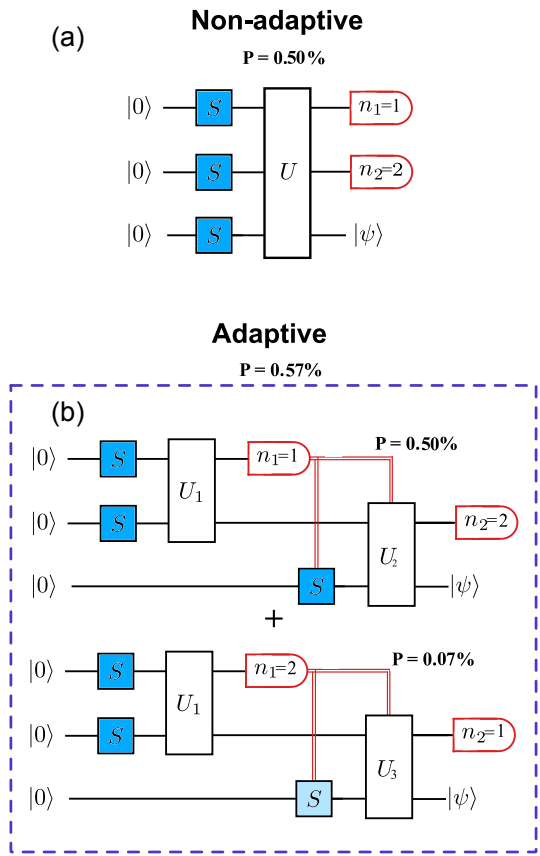} 
\caption{\label{fig:OddCatState3modes} Comparison between the (a) non-adaptive and (b) adaptive schemes for generating the odd cat state. The squeezing is bounded by $r_{\text{max}}=0.5$. The darker shades in the squeezing boxes correspond to higher squeezing intensities, with the darkest blue representing $r=0.5$. The target state is the odd cat state with $\alpha=\sqrt{6}$. The plus sign indicates that the probabilities corresponding to the two possible outcomes of the adaptive sources can be summed.} 
\end{figure}

Then, we consider an alternative optimization of the adaptive scheme, which is shown with the non-adaptive counterpart in Fig.~\ref{fig:OddCatState3modes}. In this second case, we neglect the circuit with only two modes and explore the squeezing and interferometers $U_1$ and $U_2$ that lead to the generation of states with good probability. We find that the states are generated with the same fidelity and a probability of $0.50\%$ when one photon is measured in the first mode and two photons in the second mode. Interestingly, this probability is the same as that obtained with a general interferometer acting on three modes. However, using this second scheme allows for the adaptation of the second interferometer when the measurement on the first mode does not yield the expected outcome. Specifically, if we measure two photons in the first mode, and one in the second mode, a cat state can still be generated with the same fidelity and a probability of $0.07\%$. By summing the probabilities of the adaptive schemes, we conclude that it provides an advantage over the non-adaptive one.

We observe also that if the squeezing is not bounded then choosing squeezed cat states as alternative states does not help because the complexity depends only on the core of the stellar representation of the states, and not on the total squeezing that can be factorized out.
The results for probability and fidelity obtained in this scenario are summarized in Table~\ref{tab:SqueOddCatloss}. Similarly, the results achieved by unbinding the squeezing intensity and targeting the squeezed odd cat state of amplitude $\alpha=2$ and $r=0.5$ are given in Table~\ref{tab:SqNoLimitLoss}.

\subsubsection{GKP states}\label{section:GKP state}

\begin{figure}[t] 
\centering 
\includegraphics[trim=400 280 60 285, clip,width=1\linewidth]{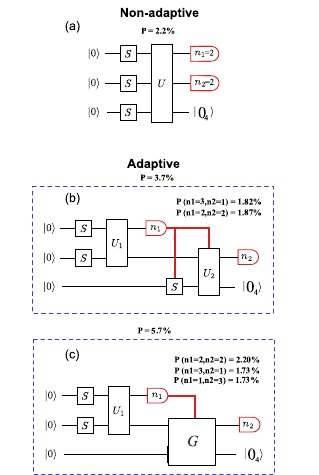} 
\caption{\label{fig:GKP4_4photonsnobounds} Comparison between (a) non-adaptive and (b,c) adaptive schemes for generating $\ket{0_4}$. Two adaptive schemes are shown in the figure: in panel (b) the scheme relies on squeezed vacuum states and passive interferometers only; in panel (c), the adaptive scheme contains a symplectic operation labeled by $G$, which here corresponds to a sequence of a beam splitter, two inline squeezers, and another beam splitter arranged as shown in Fig.~\ref{fig:BMdecomposition}.} 

\end{figure}

We consider now the core state $\ket{0_4}$ with $n_\text{max}=4$ but with a high $\Delta=10$ dB, as described in Sec.~\ref{section:metohd}. We consider three-mode circuits since the fidelity achievable with only two modes falls significantly short of $90\%$. The schemes with their probabilities are shown in Fig.~\ref{fig:GKP4_4photonsnobounds}

In this scenario, by relaxing the constraint on maximum squeezing, we observe that we can achieve fidelities greater than $99\%$ with a probability of approximately $2.2\%$ using the non-adaptive scheme (by measuring two photons in each PNRD). With the adaptive scheme, the probability increases to over $3.7\%$ (specifically, $1.8\%$ for measurement pattern (3,1) and $1.9\%$ for measurement pattern (2,2), while no parameters have been found to herald the state using other measurement patterns). In this specific case, a state with fidelity larger than $90\%$ is obtained with the same circuit optimized for the measurement pattern $(3,1)$ even when we measure $(1,3)$ which happens in $1.6\%$ of the cases. 

For even more energetic states or complex setups, we can consider schemes in which inline squeezing can be implemented in the heralded modes during intermediate steps of the scheme. Hence, the comparison involves active circuits in this context. When considering this approach, we find that the probability of generating $\ket{0_4}$ with the adaptive method is approximately $5.7\%$, as opposed to the previously reported $2.2\%$ probability for the non-adaptive method. The results for panels (a) and (c) of Fig.~\ref{fig:GKP4_4photonsnobounds} are reported in Table~\ref{tab:GKPloss_withaprioriknowledge}.

\subsubsection{Feed-forward concatenated with inline squeezing}
\begin{figure}[t]
\centering
\includegraphics[trim=310 20 200 685, clip,,width=0.9\linewidth]{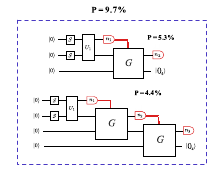}
\caption{\label{fig:concatenateGKP} Concatenated scheme to generate $\ket{0_4}$. The squares labelled by $G$ represent symplectic operations in the input mode. Here, they correspond to a sequence of a beam splitter, two inline squeezers, and another beam splitter.}
\end{figure}

As a final consideration, let's explore what happens when we extend this approach to more modes in a concatenated approach as shown in Fig.~\ref{fig:concatenateGKP}. First, if no photons are measured in the initial detectors, we can replicate the setup as it is. The results obtained for three-mode sources also hold for four-mode architectures, as smaller circuits can always be interpreted as special cases of larger circuits. We begin by considering the case in which no photons are measured in the initial detector. In this scenario, the conditional probability for the source to generate the state with these measurements is given by the product of the probability of measuring no photons in the first mode ($31\%$) and the probability of producing the state with three modes ($5.29\%$), which supplements the probability we calculated when measuring only three modes.

Similarly, looking at the other combinations when measuring up to four photons in three output modes we increase the probability of producing the state by an additional $2.77\%$.
The overall probability for this setup is then $5.29\%+(31\%\times5.29\%)+2.77\% \simeq 9.7\%$.

\subsection{Adaptive state engineering with squeezed and Fock states and PNRD}\label{section:ASE with squeezed+fock states}

\begin{figure}[h] 
\centering 
\includegraphics[trim=400 550 60 50, clip,,width=1\linewidth]{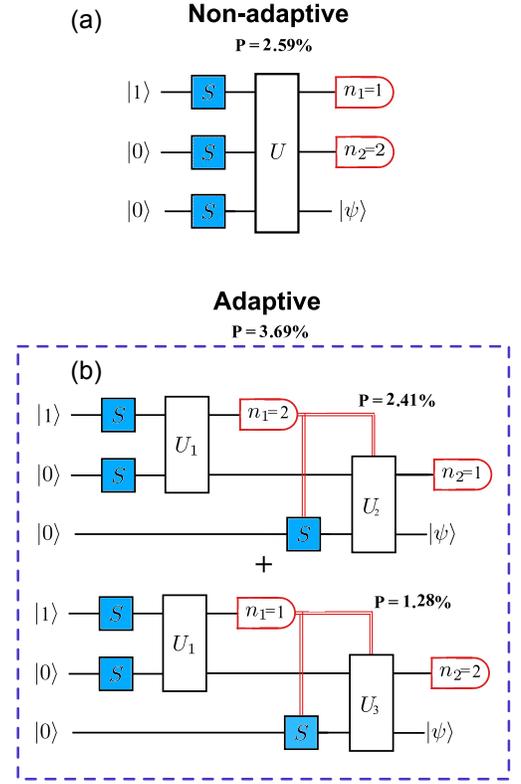} 
\caption{\label{fig:OddCatStateFock} Comparison between the (a) non-adaptive and (b) adaptive schemes for generating the odd cat state with one single photon Fock state in input. The squeezing is bounded by $r_{\text{max}}=0.5$. The darker shades in the squeezing boxes correspond to higher squeezing intensities, with the darkest blue representing $r=0.5$. The target state is the even cat state with $\alpha=\sqrt{8}$. The plus sign indicates that the probabilities corresponding to the two possible outcomes of the adaptive sources can be summed.}
\end{figure}

In this section, we evaluate whether the adaptive scheme proves advantageous even when a single-photon Fock state is used in one input mode of the circuit. The comparison between adaptive and non-adaptive schemes in this case is displayed in Fig.~\ref{fig:OddCatStateFock}. To do so, we take the even cat state with $\alpha=\sqrt{8}$ as the target of the source. 
The fidelity achieved with only two modes and measuring three photons in the output is less than $93\%$.  

Conversely, in a three-mode circuit with two two-mode interferometers $U_1$ and $U_2$  a fidelity of $95\%$ can be reached with a probability of $2.4\%$ for the measurement of one and two photons in the output ports and a probability of $1.3\%$ for a circuit where the last squeezing and $U_2$ are optimized for the measurement of two and one photon.

The overall probability for this adaptive scheme turns out to be better than the probability achievable with a non-adaptive scheme with three modes, which is equal to $2.6\%$, once again proving the advantage of an adaptive scheme in certain scenarios while maintaining the same fidelity.

\subsection{Loss and adaptivity}

Here, we examine the impact of loss on the efficacy of the various protocols. Specifically, we evaluate how the adaptive schemes behave in the presence of loss compared to the schemes where no information is fed forward. In this context, the photon loss channel can be read as a beam splitter with tunable transmission, where 100$\%$ transmission represents no loss and 0$\%$  corresponds to total loss. Unlike the setup used in~\cite{PhysRevA.101.032315}, we introduce the loss channel only after the squeezing gates at each mode, and just before detection. Here, we consider all the loss channels to have the same transmissivity. In general, when considering different losses in the output, we observe that the loss introduced in the herald modes is mainly responsible for the drop in probability, while the loss in the undetected mode affects the fidelity of the generated state.
If the loss is homogeneous among the modes, then the loss channels applied immediately after the squeezing operation in each mode can be replaced by loss channels with the same transmittivity applied after the passive interferometer. This allows us to shift all the loss introduced in our simulation to just before the measurement layer of the circuits. 
We also extend this approach to circuits relying on symplectic operations, for which the introduction of loss into all the input squeezed vacuum states cannot be done in the same way as in the case of passive interferometers alone, given that some vacuum states in the input modes interfere with other modes before being subjected to inline squeezing. 

To assess the robustness of the protocols, we initially consider an ideal scheme optimized for a specific target state, then we evaluate the probability of measuring the expected pattern and fidelity with the target under varying loss conditions. 

First, we consider the effect of loss on the sources shown in Fig~\ref{fig:OddCatState3modes} and in Fig.~\ref{fig:GKP4_4photonsnobounds}. 
In Table~\ref{tab:SqueOddCatloss}, we report the results obtained with and without loss for the two schemes depicted in Fig.~\ref{fig:OddCatState3modes}.  The target is the odd cat state with amplitude $\alpha=\sqrt{6}$ and squeezing $r =0.5$. We limit the maximum squeezing intensity to $r_{\text{max}}=0.5$.  

\begin{table}[h] 

\centering 

\begin{tabular}{ |c|c|c|c|c|c| }  

\hline 

& MP & $P$ & $P_{L=10\%}$ &$\mathcal{F}$  & $\mathcal{F}_{L=10\%}$\\ 

\hline 

Non-adaptive & 1,2 & $0.51\%$ & $0.44\%$ & $97.6\%$  & $66.1\%$\\ 

\hline 

\multirow{2}{4em}{Adaptive} & 1,2 & $0.50\%$ &$0.42\%$  & $97.6\%$  & $41.7\%$\\  

& 2,1 & $0.07\%$ &$0.30\%$& $97.6\%$  & $10.4\%$\\  

\hline 

\end{tabular} 

\caption{\label{tab:SqueOddCatloss} Results with and without $10\%$ loss for the scheme shown in Fig.~\ref{fig:OddCatState3modes}. The target is the odd cat state with amplitude $\alpha=\sqrt{6}$ squeezing $r =0.5$. We limit the maximum squeezing intensity to $r_{\text{max}}=0.5$.} 

\end{table} 

Analogously, in Table~\ref{tab:SqNoLimitLoss}, we report the results obtained without any maximum threshold on the squeezing and targeting the odd cat state with amplitude $\alpha=2$ squeezing $r =0.5$. 

\begin{center} 

\begin{table}[h] 

\centering 

\begin{tabular}{ |c|c|c|c|c|c| }  

\hline 

& MP & $P$ & $P_{L=10\%}$ &$\mathcal{F}$  & $\mathcal{F}_{L=10\%}$ \\ 

\hline 

Non-adaptive & 1,2 & $5.8\%$ & $5.54\%$ & $99.4\%$  & $69.0\%$\\ 

\hline 

\multirow{2}{4em}{Adaptive} & 1,2 & $5.8\%$ &$5.50\%$  & $99.4\%$  & $69.1\%$\\  

& 2,1 & $0.54\%$ &$1.00\%$ & $99.4\%$  & $40.7\%$\\  

\hline 

\end{tabular} 

\caption{\label{tab:SqNoLimitLoss}Results with and without $10\%$ loss for the scheme shown in Fig.~\ref{fig:OddCatState3modes}. The target is the odd cat state with amplitude $\alpha=2$ squeezing $r =0.5$.} 

\end{table}

\end{center} 

Both the results in Table~\ref{tab:SqueOddCatloss} and Table~\ref{tab:SqNoLimitLoss} show that there is no advantage in using the adaptive scheme, particularly because the fidelity achieved by the state heralded in a lossy circuit with a measurement pattern that is not used to optimize the first part of the circuit is significantly smaller than the other fidelities. A possible interpretation of this result may be connected to the fact that the optimization of the second part of the circuit is somehow deceived by a measurement that no longer corresponds to the expected heralded state. 

Now, we want to evaluate how the two schemes compare when optimized with \textit{a priori} knowledge of the loss in each mode. In this case, we consider a scheme with symplectic operations, as depicted in Fig.~\ref{fig:GKP4_4photonsnobounds}. In Table~\ref{tab:GKPloss_withaprioriknowledge}, we report the results obtained by optimizing the circuits with a loss channel defined at the end of each mode. The reported results correspond to circuits with loss levels of $L=0\%$, $L=5\%$, and $L=10\%$. Our target is the GKP state $\ket{0_4}$ with a $\Delta=10$~dB as introduced in Sec.~\ref{section:GKP state}.

\begin{center} 

\begin{table}[h] 

\centering 

\begin{tabular}{|c|c|c|c|c|c|c|c|}  

\hline 

& MP & $P$ & $P_{L=5\%}$ & $P_{L=10\%}$ &$\mathcal{F}$  & $\mathcal{F}_{L=5\%}$ & $\mathcal{F}_{L=10\%}$\\ 

\hline 

N.A. & 2,2 & $2.2\%$ & $0.11\%$& $0.055\%$ & $99.97\%$  & $85\%$& $76\%$\\ 

\hline 

\multirow{3}{2em}{Ad.} & 2,2 & $2.2\%$ &$0.048\%$ &$0.031\%$ & $99.97\%$  & $86\%$& $76\%$\\  

& 1,3 & $1.7\%$ &$0.120\%$ &$0.074\%$ & $>99.99\%$  & $82\%$& $69\%$\\  

& 3,1 & $1.7\%$ &$0.019\%$ &$0.013\%$ & $>99.99\%$  & $83\%$& $73\%$\\  

\hline 

\end{tabular} 

\caption{\label{tab:GKPloss_withaprioriknowledge} Results with and without a $5\%$ and $10\%$  loss for adaptive and non-adaptive schemes as shown in panels (a) and (c) of Fig.~\ref{fig:GKP4_4photonsnobounds}. The target state is the GKP state.} 

\end{table}

\end{center} 

As seen in Tables~\ref{tab:SqueOddCatloss}-\ref{tab:GKPloss_withaprioriknowledge} we observe smaller fidelities for the states generated with measurement pattern different from (2,2) that is the measurement pattern used to optimize the first part of the circuit. 

However, in the favorable case that even states with the smallest probability can be used in FTQC applications, the overall probability achievable with the adaptive scheme is greater than that of the non-adaptive scheme.  

  
Finally, in Table~\ref{tab:GKPloss}, we compare the results in the presence of $1\%$ when the circuit is optimized with and without the knowledge of the loss.

\begin{center} 

\begin{table*}[hbt] 

\centering 

\begin{tabular}{ |c|c|c|>{\columncolor{blue!5}}c|>{\columncolor{blue!15}}c|c|>{\columncolor{blue!5}}c|>{\columncolor{blue!15}}c| }  

\hline 

& MP & $P$ & $P_{L=1\%}$ & $P_{L=1\%}$ &$\mathcal{F}$  & $\mathcal{F}_{L=1\%}$ & $\mathcal{F}_{L=1\%}$\\ 

\hline 

Non-Adaptive & 2,2 & $2.2\%$ & $2.1\%$& $1.9\%$ & $99.97\%$  & $91.87\%$& $92.42\%$\\ 

\hline 

\multirow{3}{4em}{Adaptive} & 2,2 & $2.2\%$ &$2.1\%$ &$1.8\%$ & $99.97\%$  & $91.86\%$& $92.46\%$\\  

& 1,3 & $1.7\%$ &$1.7\%$ &$0.33\%$ & $>99.99\%$  & $88.51\%$& $90.85\%$\\  

& 3,1 & $1.7\%$ &$1.7\%$ &$1.17\%$ & $>99.99\%$  & $89.41\%$& $90.46\%$\\  

\hline 

\end{tabular} 

\caption{\label{tab:GKPloss} Results with and without a $1\%$ loss for adaptive and non-adaptive schemes as shown in panels (a) and (c) of Fig.~\ref{fig:GKP4_4photonsnobounds}. The darker columns correspond to the circuits optimized for the presence of the loss.} 

\end{table*} 
\end{center}

\section{Practical Considerations}\label{section:experiments}

The adaptive schemes we have outlined here necessitate photonic hardware with several key characteristics. Among these are high-gain squeezing, low-loss and phase-stable circuitry, optical delay lines, PNRDs and homodyne-detection. 

In free-space implementations, high-gain squeezed light generation can be achieved with second-order non-linearities in waveguide optical parametric amplifiers  (OPAs) defined in periodically-poled lithium niobate (ppLN) or potassium titanyl phosphate (ppKTP). Up to \(\sim\)20~dB of squeezing is attainable provided the losses between the squeezer and the circuit are mitigated. Recently, \(\sim\)8~dB of squeezing was measured from PPLN \cite{Kashiwazaki-8db}. Further improvements in the fabrication of these devices, and improvements in collection efficiency, should push these values towards the threshold required for GKP error correction~\cite{Tzitrin-threshold}. Electro-optic modulators (EOM) have found utility in time-bin encoded programmable interferometers~\cite{yu2024neumann}, including those used for implementing GBS \cite{Yu_2023_drug_discovery,Borealis}. Optical fibre delay lines can store the quantum states of light in a subset of modes while the remaining modes are measured and their outcomes fed forward to additional stages of linear optical networks. In bulk optics, the length of the delay lines is predicated on the switching speeds of the tuneable elements found in the adapted unitary operations in later stages. For time-bin implementations, this switching speed is typically around 1~MHz, which requires optical fibre delays of \(\sim\)200~m. 

Integrated photonics provides several attractive features, making it particularly well suited to adaptive non-Gaussian state generation. In thin-film lithium niobate (TFLN) integrated photonics, the OPA and circuit can be integrated together, thus minimizing the loss in-between. TFLN benefits from a strong electro-optic effect, enabling the integration of fast EOMs with up to 100~GHz bandwidth to facilitate fast reconfiguration of the adapted circuits. Such EOM performance should also facilitate multiplexing of multiple sources to improve further the probability of generating the state of choice. Long delay lines have recently been reported in time-bin entanglement experiments on TFLN \cite{finco2024time}. Without integrated detectors on-chip, the out-coupling efficiency becomes a limiting factor in the end-to-end system efficiency. In TFLN, grating couplers with greater than $80\%$ coupling efficiency have been demonstrated by incorporating metal mirrors~\cite{Chen-gratingcouplers} to reduce absorption in the substrate.

Finally, PNRDs are required to herald the desired non-Gaussian state. Superconducting transition-edge sensors (TES), with operating temperatures of \(\sim\)100~mK, are currently the leading technology in this area. Detection efficiencies $>95\%$ can be achieved in the telecom C-band with photon-number resolution of up to 20 photons. TES typically suffer from long-reset times, limiting the rate at which states can be heralded to \(\sim\)100~kHz. However, \(\sim\)1~MHz rates have recently been reported~\cite{hummatov_fast_2023} by improving the heat dissipation from the TES to the substrate, which together with advances in TES signal processing~\cite{ZliTES} could allow state generation rates approaching 5~MHz.
For highly efficient detectors capable of distinguishing small photon numbers, the fidelity of the generated states is not significantly affected by imperfections. This is because the expected number of detected photons is generally low, and at least two erroneous detections, one overestimating and one underestimating the result, must occur within the same iteration to herald a state different from the target.

\section{Conclusion}\label{section:conclusion}
We introduced an alternative scheme for non-Gaussian quantum state generation based on the GBS-like source introduced in~\cite{PhysRevA.100.052301}. In our approach, rather than employing an input layer with squeezed vacuum states, a universal linear network layer, and a measurement layer, we considered a scheme consisting of several layers. Each layer has its own input states, unitary operations, and PNRDs, forming a concatenation of GBS-like non-Gaussian sources, where the output state of one layer is part of the input state for the next. The information on the number of photons detected is fed forward through the circuit, allowing the interferometer's parameters to be adapted accordingly. We used the number of modes as the key resource to ensure a fair comparison between our proposed scheme and the original. We optimized all parameters to maximize both the fidelity of the output state and the probability of success. Our results demonstrate an improvement in the quality of non-Gaussian state generation with the adaptive scheme, in terms of either success probability or fidelity. Even when a vacuum state is replaced by a single-photon state at the input, we observe a relative increase in probability of over 40$\%$, attributable to the adaptive approach. Finally, simulations conducted on lossy circuits reveal that, although losses negatively impact both adaptive and non-adaptive schemes, the adaptive scheme remains preferable. More specifically, as seen in Table~\ref{tab:GKPloss_withaprioriknowledge}, the adaptive scheme is more resilient to doubling the loss compared to the non-adaptive approach. Our work provides a more versatile framework to optimise the generation of non-Gaussian resource states for applications in photonic quantum sensing and computation.

\begin{acknowledgments}\label{section:acknowledgments}
R.B.P. would like to thank Miller Eaton for the fruitful discussions which motivated this work. R.B.P. acknowledges support from the  UK Research \& Innovation Future Leaders Fellowship program (project number: MR/W011794/1), the Engineering and Physical Sciences Research Council (EPSRC) UK Quantum Technologies Program's hubs for Quantum Computing \& Simulation (project number: EP/T001062/1) and Quantum Computing via Integrated and Interconnected Implementations (project number: EP/Z53318X/1). S.Y. is supported by the UK Research and Innovation Guarantee Postdoctoral Fellowship (project number: EP/Y029631/1). V.C., K.H. and R.B.P. acknowledge support from Mitacs Globalink Research Award (IT37329). K.H. acknowledges support from NSERC’s Discovery Grant and Alliance programs.
\end{acknowledgments}


%

\end{document}